\documentclass[12pt,a4paper]{article}
\usepackage{amsmath,amssymb,amsthm} 

\newtheorem{prop}{Proposition}
\newtheorem{cor}{Corollary}
\theoremstyle{remark}
\newtheorem{remark}{Remark}
\newcommand{\beq}{\begin{equation}}
\newcommand{\eeq}{\end{equation}}
\newcommand{\beqnn}{\begin{equation*}}
\newcommand{\eeqnn}{\end{equation*}}
\newcommand{\rd}{\partial}

\newcommand{\CC}{\mathbb{C}}
\newcommand{\PP}{\mathbb{P}}

\newcommand{\frakL}{\mathfrak{L}}
\newcommand{\bsx}{\boldsymbol{x}}
\newcommand{\bst}{\boldsymbol{t}}

\newcommand{\bszero}{\boldsymbol{0}}

\newcommand{\Wbar}{\bar{W}}

\newcommand{\tbar}{\bar{t}}

\newcommand{\calL}{\mathcal{L}}
\newcommand{\calT}{\mathcal{T}}

\begin{document}

\title{Generalized ILW hierarchy: Solutions and limit 
to extended lattice GD hierarchy}
\author{Kanehisa Takasaki\thanks{E-mail: takasaki@math.kindai.ac.jp}\\
{\normalsize Department of Mathematics, Kindai University}\\ 
{\normalsize 3-4-1 Kowakae, Higashi-Osaka, Osaka 577-8502, Japan}}
\date{}
\maketitle

\begin{abstract}
The intermediate long wave (ILW) hierarchy and its generalization,  
labelled by a positive integer $N$, can be formulated as 
reductions of the lattice KP hierarchy. The integrability of 
the lattice KP hierarchy is inherited by these reduced systems.  
In particular, all solutions can be captured by a factorization 
problem of difference operators.  A special solution among them 
is obtained from Okounkov and Pandharipande's dressing operators 
for the equivariant Gromov-Witten theory of $\mathbb{CP}^1$.  
This indicates a hidden link with the equivariant Toda hierarchy.  
The generalized ILW hierarchy is also related to the lattice 
Gelfand-Dickey (GD) hierarchy and its extension by logarithmic flows.  
The logarithmic flows can be derived from the generalized 
ILW hierarchy by a scaling limit as a parameter of the system 
tends to $0$.  This explains an origin of the logarithmic flows.  
A similar scaling limit of the equivariant Toda hierarchy yields 
the extended 1D/bigraded Toda hierarchy.    
\end{abstract}

\begin{flushleft}

\end{flushleft}

\newpage

\section{Introduction}

The intermediate long wave (ILW) equation 
is an integro-differential equation that describes 
internal long waves in a stratified fluid of finite depth. 
The KdV and Benjamin-Ono equations may be thought of 
as its limit as the depth tends to $0$ and $\infty$ 
respectively.  Soon after the equation was proposed 
\cite{Joseph77,KKD78}, many features of integrability, 
such as soliton solutions, B\"acklund transformations, 
inverse scattering transformations, conservation laws 
and Hamiltonian structures, 
were discovered \cite{SAK79,CL79,KSA81,LR83}.  
Since then, further new aspects have been reported 
on this equation, its hierarchy of higher flows 
(the ILW hierarchy) and related integrable systems 
\cite{DLOPPS91,DLOPPS92,TS03,ST09}. 

Recently, Buryak and Rossi presented 
a new Lax representation of the ILW hierarchy \cite{BR18b}. 
Their Lax representation is formulated in terms 
of difference operators just like integrable hierarchies 
of the Toda type \cite{Takasaki18}.  They compared it 
to the equivariant bigraded Toda hierarchy \cite{MT07}
and argued that the ILW hierarchy is a degenerate case 
thereof.  The bigraded equivariant Toda hierarchy, 
as well as the ordinary equivariant 1D Toda hierarchy 
\cite{Getzler04}, is a reduction of the 2D Toda hierarchy.  
An approach to the ILW hierarchy 
from the 2D Toda hierarchy was indeed carried out 
by Liu et al. \cite{LWZ21}.  The original form 
of Buryak and Rossi's Lax representation, 
however, rather resembles the lattice KP hierarchy 
(aka the discrete KP hierarchy, the modified KP hierarchy, 
etc. \cite{Dickey-book}).  

We consider the ILW hierarchy and its generalization, 
labelled by a positive integer $N$, as reductions 
of the lattice KP hierarchy.  These reduced systems 
have a parameter $\nu$.  If $\nu = 0$, 
the same reduction procedure yields 
the lattice Gelfand-Dickey (GD) hierarchy. 
In the previous work \cite{Takasaki22}, 
we constructed an extension of the lattice GD hierarchy. 
A main subject of this paper is to elucidate 
the relationship among these systems.  

The extended lattice GD hierarchy is obtained 
by adding, by hand, a set of \textit{logarithmic flows}. 
We are interested in the origin of these flows. 
A similar extension by logarithmic flows is known 
for the 1D Toda hierarchy \cite{CDZ04,Takasaki10} 
and its bigraded generalization \cite{Carlet06,Li-etal09}.  
We shall show that the $N$-th generalized ILW hierarchy 
turns into the $N$-th extended lattice GD hierarchy 
in a limit as the parameter $\nu$ tends to $0$.  
This is achieved by a kind of scaling limit 
of the time variables of flows.  In this limit, 
part of the flows of the generalized ILW hierarchy 
are transmuted into the logarithmic flows. 
Actually, we can show that the extended 1D/bigraded 
Toda hierarchy, too, can be derived from 
the equivariant 1D/bigraded Toda hierarchy  
by a similar scaling limit.  The parameter $\nu$ 
therein plays the role of equivariant parameter. 
Thus the extended Toda hierarchy turns out 
to be a non-equivariant limit of the equivariant 
Toda hierarchy.  To the best of our knowledge, 
this fact has not been explained in such a direct way. 

Another subject of this paper is the description 
of solutions of the generalized ILW hierarchy.  
We shall consider two types of solutions. 
The first one are soliton solutions.  Soliton solutions 
of the lattice KP hierarchy are well known.  
We show a condition under which they become solutions 
of the generalized ILW hierarchy.  This condition 
is described by a set of equations for the parameters 
of the soliton solutions.  We can use these equations 
to illustrate the aforementioned scaling limit 
to the extended lattice GD hierarchy.  
Solutions of the second type are characterized 
by a factorization problem of difference operators.  
The factorization problem itself can generate 
all solutions of the generalized ILW hierarchy.  
We find a special solution of this type 
in the context of our previous work \cite{Takasaki21} 
on the equivariant Toda hierarchy.  We constructed therein 
a pair of difference operators that play the role 
of Okounkov and Pandharipande's ``dressing operators''  
\cite{OP02,Johnson09} in their fermionic description 
of the equivariant Gromov-Witten theory of $\CC\PP^1$.  
One of the two operators can be used for a special setup 
of the factorization problem, which generates a solution 
of the generalized ILW hierarchy.  We show 
a generalization of this special solution as well. 

This paper is organized as follows.  
Section 2 is a review of the lattice KP and 
generalized ILW hierarchies.  Section 3 introduces 
the factorization problem in a general form.  
In Section 4, special solutions of the two types 
are presented.  In Section 5, the scaling limit 
to the extended lattice GD hierarchy is formulated 
and illustrated for soliton solutions.  
An appendix is added for a supplementary explanation 
on the bigraded equivariant Toda hierarchy 
and its scaling limit to the extended bigraded 
Toda hierarchy.

\section{Lattice KP and generalized ILW hierarchies}

The 2D Toda hierarchy has two Lax operators $L,\bar{L}$ 
and two sets of time variables $\bst = (t_k)_{k=1}^\infty$, 
$\bar{\bst} = (\tbar_k)_{k=1}^\infty$ \cite{Takasaki18}
(see Appendix).  The lattice KP hierarchy 
can be obtained from the 2D Toda hierarchy 
by discarding the second Lax operator $\bar{L}$ 
and the second set of time variables $\bar{\bst}$. 
We here consider an $\hbar$-dependent formulation 
of these integrable hierarchies \cite{TT95}.  
Let $s$ be the spatial coordinate therein 
and $\Lambda$ denote the shift operator 
\[
  \Lambda = e^{\hbar\rd_s}, \quad \rd_s = \rd/\rd s, 
\]
that acts on a function $f(s)$ of $s$ as 
$\Lambda^n f(s) = f(s + n\hbar)$.  

\subsection{Lattice KP hierarchy}

The $\hbar$-dependent lattice KP hierarchy 
consists of the Lax equations 
\beq
  \hbar\frac{\rd L}{\rd t_k} = [B_k,L],\quad k = 1,2,\ldots,
  \label{Laxeq}
\eeq
for the difference Lax operator 
\[
  L = \Lambda + \sum_{n=1}^\infty u_n\Lambda^{1-n}
\]
(which is an analogue of the pseudo-differential Lax operator 
of the KP hierarchy).   The coefficients 
$u_n = u_n(\hbar,s,\bst)$ are dynamical variables 
that depend on the space-time coordinates $s$ and 
$\bst = (t_k)_{k=1}^\infty$ and the parameter $\hbar$.  
$B_k$'s are defined by the Lax operator as 
\[
  B_k = (L^k)_{\ge 0},
\]
where $(\quad)_{\ge 0}$ is the projection 
\[
  \left(\sum_{n=-\infty}^\infty a_n\Lambda^n\right)_{\ge 0}
  = \sum_{n\ge 0}a_n\Lambda^n 
\]
to the linear combination of non-negative powers 
of $\Lambda$.  Let $(\quad)_{<0}$ denote 
the complementary projection 
\[
  \left(\sum_{n=-\infty}^\infty a_n\Lambda^n\right)_{<0}
  = \sum_{n<0}a_n\Lambda^n. 
\]

The wave function $\Psi$ of the auxiliary linear equations 
\beq
  \hbar\frac{\rd\Psi}{\rd t_k} = B_k\Psi,\quad 
  L\Psi = z\Psi 
  \label{lineq}
\eeq
is related to the tau function $\tau = \tau(\hbar,s,\bst)$ 
as 
\beq
  \Psi 
  = \frac{\tau(\hbar,s,\bst - \hbar[z^{-1}])}
    {\tau(\hbar,s,\bst)}z^{s/\hbar}e^{\xi(\bst,z)/\hbar}, 
  \label{Psi-tau}
\eeq
where 
\[
  [z] = (z^k/k)_{k=1}^\infty,\quad 
  \xi(\bst,z) = \sum_{k=1}^\infty t_kz^k. 
\]

The amplitude part of $\Psi$ can be expanded into 
negative powers of $z$ as 
\[
  \frac{\tau(\hbar,s,\bst - \hbar[z^{-1}])}
  {\tau(\hbar,s,\bst)}
  = 1 + \sum_{n=1}^\infty w_nz^{-n} 
\]
and used to construct the dressing operator 
\[
  W = 1 + \sum_{n=1}^\infty w_n\Lambda^{-n}. 
\]
The wave function, in turn, can be written as 
\[
  \Psi = Wz^{s/\hbar}e^{\xi(\bst,z)/\hbar}. 
\]
$L$ and $B_k$'s are expressed in the dressed form as 
\[
  L = W\Lambda W^{-1},\quad 
  B_k = (W\Lambda^kW^{-1})_{\ge 0}. 
\]
The Lax equations are thus converted to the Sato equations 
\beq
  \hbar\frac{\rd W}{\rd t_k} 
  = (W\Lambda^k W^{-1})_{\ge 0}W - W\Lambda^k 
  = - (W\Lambda^k W^{-1})_{<0}W 
  \label{Satoeq}
\eeq
for the dressing operator $W$.

\subsection{Generalized ILW hierarchy}

Let $N$ be a positive integer and allow 
the dynamical variables to depend on the new parameter $\nu$ 
as $u_n = u_n(\hbar,\nu,s,\bst)$, etc.  
Reduction to the $N$-th generalized ILW hierarchy 
with parameter $\nu$ is achieved 
by the reduction condition 
\beq
  (L^N - \nu\log L)_{<0} = 0. 
  \label{red-cond}
\eeq
If $\nu = 0$, this condition becomes 
the $N$-reduction condition $(L^N)_{<0} = 0$ 
to the lattice GD hierarchy. 
The notation used here in the $\nu \not= 0$ case 
will need explanation \cite{BR18b,Takasaki22}. 
First, the logarithm $\log L$ of $L$ is defined as 
\[
  \log L = W\log\Lambda W^{-1} = W\hbar\rd_s W^{-1}.
\]
Second, the projections $(\quad)_{\ge}$ and $(\quad)_{<0}$ 
of difference operators are extended to operators 
of the form  $A\rd_s + B$, $A$ and $B$ being 
genuine difference operators, as 
\[
\begin{aligned}
  (A\rd_s + B)_{\ge 0} &= (A)_{\ge 0}\rd_s + (B)_{\ge 0},\\
  (A\rd_s + B)_{<0} &= (A)_{<0}\rd_s + (B)_{<0}. 
\end{aligned}
\]
This extended projection should be used carefully. 
Namely, the shift operators $\Lambda^n = e^{n\hbar\rd_s}$ 
should not be expanded into powers of $\rd_s$; 
such careless expansion leads to ambiguity 
in the definition of the projection.  

Under the reduction condition (\ref{red-cond}), 
the lattice KP hierarchy reduces to a system 
of evolution equations for a finite number of 
dynamical variables.  Note that $L^N - \nu\log L$ 
can be expressed as 
\[
  L^N - \nu\log L 
  = B_N - \nu\log\Lambda + \text{negative powers of $\Lambda$}. 
\]
The reduction condition implies that 
the ``negative powers of $\Lambda$'' on the right hand side 
disappear.  We thus obtain the reduced Lax operator 
\beq
  \frakL = L^N - \nu\log L = B_N - \nu\log\Lambda 
  \label{red-Laxop}
\eeq
that satisfies the reduced Lax equations
\beq
  \hbar\frac{\rd\frakL}{\rd t_k} = [B_k,\frakL], \quad 
  k = 1,2,\ldots. 
  \label{red-Laxeq}
\eeq
$B_N$ is a difference operator of the form 
\[
  Q = \Lambda^N + b_1\Lambda^{N-1} + \cdots + b_N, 
\]
and $B_k$'s can be thereby redefined as 
\[
  B_k = (Q^{k/N})_{\ge 0}. 
\]
Fractional powers and logarithm of $Q$ 
\cite{CDZ04,Carlet06,LWZ21} can be used to reconstruct 
$L$ and $\log L$ as well.  Thus the Lax equations 
(\ref{red-Laxeq}) become a system of evolution equations 
for $b_n$'s. 

The reduction condition (\ref{red-cond}) can be translated 
to the language of the dressing operator 
and the tau function as follows.  

\begin{prop}
The reduction condition (\ref{red-cond}) is equivalent 
to the equation 
\beq
  \frac{\rd W}{\rd t_N} - \nu\frac{\rd W}{\rd s} = 0  
  \label{W-red-cond}
\eeq
for the dressing operator. 
\end{prop}

\begin{proof}
The condition (\ref{red-cond}) implies that 
\beq
  W(\Lambda^N - \nu\log\Lambda)W^{-1} = B_N - \nu\log\Lambda, 
  \label{W-B}
\eeq
hence 
\[
  W(\Lambda^N - \nu\log\Lambda) = (B_N - \nu\log\Lambda)W. 
\]
Since 
\[
  B_NW = \hbar\frac{\rd W}{\rd t_N} + W\Lambda^N,\quad 
  \log\Lambda W = \hbar\frac{\rd W}{\rd s} + W\log\Lambda, 
\]
the last relation boils down to (\ref{W-red-cond}).  
\end{proof}

\begin{cor}
If the tau function of the lattice KP hierarchy 
satisfies the equation 
\beq
  \frac{\rd\tau}{\rd t_N} - \nu\frac{\rd\tau}{\rd s} = 0, 
  \label{tau-red-cond}
\eeq
the associated Lax and dressing operators satisfy 
the reduction conditions (\ref{red-cond}) 
and (\ref{W-red-cond}) to the generalized ILW hierarchy. 
\end{cor}

The reduction conditions (\ref{W-red-cond}) and 
(\ref{tau-red-cond}) mean that the tau function 
$\tau = \tau(\hbar,\nu,s,\bst)$ and all dynamical variables 
of the lattice KP hierarchy depend on $t_N$ and $s$ 
through the linear combination $t_N + s/\nu$. 
In particular, the tau function can be expressed as 
\beq
  \tau = f(\hbar,\nu,t_1,\ldots,t_{N-1},t_N + s/\nu,t_{N+1},\ldots) 
  \label{red-tau}
\eeq
with some function $f(\hbar,\nu,\bst)$ of $\hbar$, $\nu$ and $\bst$. 

\begin{remark}
The Lax equations (\ref{red-Laxeq}) define 
isospectral deformations of the spectral problem 
\beq
  \frakL\Psi = \lambda\Psi. 
  \label{fkL-lineq}
\eeq
The spectral variable $\lambda$ is related 
to the spectral parameter $z$ as 
\beq
  \lambda = z^N - \nu\log z. 
  \label{lam-z}
\eeq
We shall encounter avatars of this relation 
in the construction of soliton solutions. 
\end{remark}

\subsection{Relation to ILW equation}

The ILW equation \cite{Joseph77,KKD78} is related 
to the $N = 1$ case.  The reduced Lax operator (\ref{red-Laxop}) 
in this case takes the form 
\[
  \frakL = \Lambda + u - \nu\log\Lambda, \quad 
  u = u_1, 
\]
proposed by Buryak and Rossi \cite{BR18b}.  
The Lax equations (\ref{red-Laxeq}) for $k = 1,2$ yield 
the equations 
\begin{align}
  \hbar\frac{\rd u}{\rd t_1} &= \nu\hbar\frac{\rd u}{\rd s},
  \label{u-t1eq}\\
  \hbar\frac{\rd u}{\rd t_2} &= \nu\hbar\frac{\rd b_{22}}{\rd s} 
  \label{u-t2eq}
\end{align}
for $u$, where $b_{22}$ denotes the last term of 
\[
  B_2 = \Lambda^2 + b_{12}\Lambda + b_{22}. 
\]
Actually, $b_{22}$ can be expressed in terms of $u$ 
as follows.  

\begin{prop}
\beq
  b_{22} = u^2 + \nu(1 + \Lambda)(1 - \Lambda)^{-1}\hbar\frac{\rd u}{\rd s}. 
  \label{b22}
\eeq
\end{prop}

\proof 
Since $L^2$ can be expanded into powers of $\Lambda$ as 
\[
  L^2 = \Lambda^2 + (u_1(s) + u_1(s+\hbar))\Lambda 
        + (u_1(s)^2 + u_2(s) + u_2(s+\hbar)) + \cdots, 
\]
$b_{22}$ can be written as 
\[
  b_{22} = u^2 + (1 + \Lambda)u_2. 
\]
To find an expression of $u_2$, let us recall 
the relation (\ref{red-Laxop}) between $L$ and $\frakL$: 
\[
  L - \nu\log L = \frakL = \Lambda + u - \nu\log\Lambda.
\]
Since 
\[
  \log L 
  = W\Lambda W^{-1} 
  = \log\Lambda - \hbar\frac{\rd W}{\rd s}W^{-1} 
  = \log\Lambda - \hbar\frac{\rd w_1}{\rd s}\Lambda^{-1} + \cdots,  
\]
the left hand side of the last relation is an operator 
of the form 
\[
  L - \nu\log L 
  = - \nu\log\Lambda + \Lambda + u_1 
    + \left(u_2 - \nu\hbar\frac{\rd w_1}{\rd s}\right)\Lambda^{-1} 
    + \cdots.
\]
There is, however, no $\Lambda^{-1}$-term 
in $\frakL = \Lambda + u - \nu\log\Lambda$. 
Consequently, we find that 
\[
  u_2 = \nu\hbar\frac{\rd w_1}{\rd s}, 
\]
so that 
\[
  b_{22} = u^2 + \nu(1 + \Lambda)\hbar\frac{\rd w_1}{\rd s}. 
\]
To find an expression of $w_1$, let us notice 
that the $\Lambda^0$-part of the intertwining relation 
$W\Lambda = LW$ reads 
\[
  w_1(s) = u_1(s) + w_1(s+\hbar). 
\]
This implies that 
\[
  (1 - \Lambda)w_1 = u_1.
\]
Solving this relation for $w_1$ as 
\[
  w_1 = (1 - \Lambda)^{-1}u
\]
and plugging it into the foregoing expression 
of $b_{22}$, we obtain (\ref{b22}). 
\qed

By (\ref{b22}), the equation (\ref{u-t2eq}) 
of the $t_2$-flow becomes a closed equation for $u$: 
\beq
  \hbar\frac{\rd u}{\rd t_2} 
  = \nu\hbar\frac{\rd}{\rd s}\left(u^2 
    + \nu(1 + \Lambda)(1 - \Lambda)^{-1}\hbar\frac{\rd u}{\rd s}\right). 
  \label{u-t2eq-bis}
\eeq
If $\hbar$ takes an imaginary value $\hbar = 2i\delta$, $\delta > 0$, 
the non-local operator $(1 + \Lambda)(1 - \Lambda)^{-1}$ 
can be identified with ($i$ times) the integral operator 
\cite{KSA81,LR83}
\[
  T[u](x) = \rm{p.v.}\int_{-\infty}^\infty \frac{1}{2\delta}
     \coth\left(\frac{y-x}{2\delta}\right)u(y)dy
\]
in the usual formulation 
\[
  \frac{\rd u}{\rd t} + \frac{1}{\delta}\frac{\rd u}{\rd x} 
  + 2u\frac{\rd u}{\rd x} 
  + T\left[\frac{\rd^2 u}{\rd x^2}\right] = 0 
\]
of the ILW equation.  Taking a linear combination 
with the equation (\ref{u-t1eq}) of the $t_1$-flow 
amounts to adding the advection term $\rd u/\rd x$. 
Thus the $N = 1$ case of the generalized ILW hierarchy 
turns out to contain the ILW equation in the lowest part.

\section{Factorization problem}

\subsection{Factorization problem for lattice KP hierarchy}

We now turn to the issue of solving the generalized 
ILW hierarchy by a factorization problem 
of difference operators.  

To this end, let us start from the factorization problem 
\beq
  U = W^{-1}\bar{W}
  \label{fac-prob}. 
\eeq
for the lattice KP hierarchy.  
The first factor on the right hand side 
is the inverse of a difference operator $W$ 
of the type considered in the previous section. 
The second factor $\bar{W}$ is a difference operator 
\[
  \bar{W} = \sum_{n=0}^\infty\bar{w}_n\Lambda^n,\quad
  \bar{w}_0 \not=0, 
\]
of the opposite type.  The solution-generating operator $U$ 
is assumed to satisfy the simple evolution equations 
\beq
  \hbar\frac{\rd U}{\rd t_k} = \Lambda^k U, \quad 
  k = 1,2,\ldots, 
  \label{U-lineq}
\eeq
and the problem is to factorize it as shown in (\ref{fac-prob}). 

One can confirm, as illustrated in our previous work 
\cite{Takasaki22}, that the first factor $W$ 
of the solution $(W,\bar{W})$ of this factorization problem 
does satisfy the Sato equations (\ref{Satoeq}).  
Moreover, as explained in the remarks below, all solutions 
of the lattice KP hierarchy can be thus captured. 

To fulfill the reduction condition (\ref{red-cond}) 
to obtain a solution of the generalized ILW hierarchy, 
one needs an extra condition on $U$.  
We shall present a necessary and sufficient condition 
in the next subsection. 

\begin{remark}
The second factor $\bar{W}$, which is absent 
in the formulation of the lattice KP hierarchy, 
emerges from the Sato equations (\ref{Satoeq}) directly 
as follows.  One can rewrite these equations as 
\[
  \hbar\frac{\rd}{\rd t_k}(We^{\xi(\bst,\Lambda)/\hbar}) 
  = B_kWe^{\xi(\bst,\Lambda)/\hbar}. 
\]
Let $\bar{W}$ be the difference operator 
\beq
  \bar{W} = We^{\xi(\bst,\Lambda)/\hbar}W_{\rm{in}}^{-1},
  \label{Wbar-W0}
\eeq
where $W_{\rm{in}}$ denotes the initial value 
\[
    W_{\rm{in}} = W|_{\bst=\bszero} 
\]
of $W$ at $\bst = \bszero$.  $\bar{W}$ satisfies 
the evolution equations 
\beq
  \hbar\frac{\rd\bar{W}}{\rd t_k} = B_k\bar{W} 
  \label{Wbar-Satoeq}
\eeq
and the initial condition 
\[
  \bar{W}|_{\bst=\bszero} = 1.
\]
This implies that all Taylor coefficients 
of $\bar{W}$ at $\bst = \bszero$ contain 
no negative powers of $\Lambda$.  
Thus $\bar{W}$ turns out to be a difference operator 
of the aforementioned form.  
One can rewrite (\ref{Wbar-W0}) as 
\beq
  e^{\xi(\bst,\Lambda)/\hbar}W_{\rm{in}}^{-1} = W^{-1}\bar{W}.  
  \label{W0-fac-prob}
\eeq
Since the operator on the left hand side satisfies 
the differential equations (\ref{U-lineq}), this is exactly 
a factorization relation of the form shown in (\ref{fac-prob}). 
\end{remark}

\begin{remark}
(\ref{W0-fac-prob}) is a formulation 
of the factorization problem that solves 
the Sato equations (\ref{Satoeq}) under 
the initial condition $W|_{\bst=\bszero} = W_{\rm{in}}$. 
Moreover, $W$ does not change even if 
this special choice of $U$ is modified to 
\beq
  U = e^{\xi(\bst,\Lambda)/\hbar}W_{\rm{in}}^{-1}C, 
  \label{U-W0}
\eeq
where $C$ is an arbitrary difference operator 
of the form 
\[
  C = \sum_{n=0}^\infty c_n\Lambda^n,\quad c_0 \not= 0, 
  \quad \frac{\rd C}{\rd t_k} = 0, \quad k=1,2,\ldots. 
\]
In the context of the initial value problem, 
(\ref{U-W0}) is the most general choice of $U$.  
\end{remark}

\subsection{Reduction condition on $U$-operator}

We now allow $U$ to depend on $\nu$ as well, 
and seek for a condition on $U$ under which 
the solution of the factorization problem (\ref{fac-prob}) 
yields a solution of the generalized ILW hierarchy. 
Such a condition can be formulated as follows. 

\begin{prop}
Let $L$ be the Lax operator $L = W\Lambda W^{-1}$ 
of the lattice KP hierarchy obtained from 
the factorization problem (\ref{fac-prob}). 
$L$ satisfies the reduction condition (\ref{red-cond}) 
to the generalized ILW hierarchy if and only if 
there is a difference operator 
\beq
  \varphi = \sum_{n=0}^\infty\varphi_n\Lambda^n,\quad 
  \frac{\rd\varphi}{\rd t_k} = 0,\quad k = 1,2,\ldots, 
  \label{varphi}
\eeq
such that $U$ satisfies the intertwining relation
\beq
  (\Lambda^N - \nu\log\Lambda)U = U(\varphi - \nu\log\Lambda). 
  \label{U-red-cond}
\eeq
\end{prop}

\begin{proof}
Suppose that $L$ satisfies the reduction condition 
(\ref{red-cond}). Let us recall the algebraic relation 
(\ref{W-B}) mentioned in the proof of (\ref{W-red-cond}). 
Substituting $W = \bar{W}U^{-1}$ and $W^{-1} = U\bar{W}^{-1}$ 
therein yields the equation 
\[
  U^{-1}(\Lambda^N - \nu\log\Lambda)U 
  = \bar{W}^{-1}(B_N - \nu\log\Lambda)\bar{W}. 
\]
Since 
\[
  \bar{W}^{-1}\log\Lambda\bar{W} 
  = \bar{W}^{-1}\hbar\frac{\rd\bar{E}}{\rd s} + \log\Lambda, 
\]
we can rewrite the right hand side of the last equation as 
\[
  \bar{W}^{-1}(B_N - \nu\log\Lambda)\bar{W} 
  = \bar{W}^{-1}B_N\bar{W} 
   - \bar{W}^{-1}\nu\hbar\frac{\rd\bar{W}}{\rd s} 
    - \nu\log\Lambda. 
\]
Thus (\ref{U-red-cond}) holds by letting 
\beq
  \varphi 
  = \bar{W}^{-1}B_N\bar{W} - \bar{W}^{-1}\nu\hbar\frac{\rd\bar{W}}{\rd s}. 
  \label{varphi-BWbar}
\eeq
Obviously, $\varphi$ is a difference operator 
without negative powers of $\Lambda$.  Moreover, 
differentiating both sides of (\ref{U-red-cond}) 
with respect to $t_k$ and using (\ref{U-lineq}), 
we obtain the equation 
\[
  (\Lambda^N - \nu\log\Lambda)\Lambda^kU 
  = \Lambda^kU(\varphi - \nu\log\Lambda) 
    + U\hbar\frac{\rd\varphi}{\rd t_k}. 
\]
Since the left hand side and the first term 
on the right hand side cancel out 
by (\ref{U-red-cond}) itself, we can conclude 
that $\rd\varphi/\rd t_k = 0$.  

Conversely, suppose that there is a difference operator 
$\varphi$ as shown in (\ref{varphi}) 
such that (\ref{U-red-cond}) holds.  
(\ref{U-red-cond}) turns into the equation 
\beq
  W(\Lambda^N - \nu\log\Lambda)W^{-1} 
  = \bar{W}(\varphi - \nu\log\Lambda)\bar{W}^{-1} 
  \label{WWbar-rel}
\eeq
for $W$ and $\bar{W}$.  We can further rewrite 
this equation as 
\[
  L^N - \nu\log L
  = \bar{W}\varphi\bar{W}^{-1} 
    + \nu\hbar\frac{\rd\bar{W}}{\rd s}\bar{W}^{-1} 
    - \nu\log\Lambda. 
\]
Since the first and second terms on the right hand side 
contain no negative powers of $\Lambda$, (\ref{red-cond}) holds.  
\end{proof}

\begin{remark}
If $\nu = 0$, (\ref{U-red-cond}) reduces 
to the reduction condition to the lattice GD hierarchy 
in our previous work \cite{Takasaki22}. 
\end{remark}

\begin{remark}
(\ref{U-red-cond}) is equivalent to the intertwining relation 
\beq
  (\Lambda^N - \nu\log\Lambda)U_{\rm{in}} 
  = U_{\rm{in}}(\varphi - \nu\log\Lambda)
  \label{Uin-red-cond}
\eeq
for the initial value $U_{\rm{in}} = U|_{\bst=\bszero}$. 
\end{remark}

\begin{remark}
By the Sato equations (\ref{Wbar-Satoeq}) for $\Wbar$, 
one can rewrite (\ref{varphi-BWbar}) as 
\[
 \varphi = \bar{W}^{-1}\hbar\left(\frac{\rd\bar{W}}{\rd t_N} 
            - \nu\frac{\rd\bar{W}}{\rd s}\right). 
\]
Thus the dependence of $\Wbar$ on $t_N$ and $s$ 
is not as simple as that of $W$, cf. (\ref{W-red-cond}). 
\end{remark}

\begin{remark}
The coefficients $\varphi_n$ of $\varphi$ can depend $s$. 
If they are constants, (\ref{WWbar-rel}) becomes 
the algebraic relation 
\beq
  L^N - \nu\log L = \sum_{n=0}^\infty\varphi_n\bar{L}^n - \nu\log\bar{L}
  \label{LLbar-rel}
\eeq
between $L$ and 
\[
  \bar{L} = \bar{W}\Lambda\bar{W}^{-1}. 
\]
$\bar{L}$ amounts to the second Lax operator 
of the 2D Toda hierarchy, 
\[
  \bar{L}^{-1} = \sum_{n=0}^\infty\bar{u}_n\Lambda^{n-1}, 
\]
though it plays no dynamical role 
in the present setup where the second time variables 
$\bar{\bst} = (\bar{t}_k)_{k=1}^\infty$ 
of the 2D Toda hierarchy are turned off. 
The algebraic relation (\ref{LLbar-rel}) 
in the $\varphi = 0$ (and $N = 1$) case 
coincides with the reduction condition 
proposed by Liu et al. \cite{LWZ21} to characterize 
the ILW hierarchy in the 2D Toda hierarchy. 
Actually, as far as $\varphi_n$'s are constants, 
$\varphi$ can be eliminated by a reparametrization transformation 
\[
  \bar{L} \to c_0\bar{L} + c_1\bar{L}^2 + \cdots, \quad 
  c_0 \not= 0, 
\]
of $\bar{L}$ with constant coefficients $c_n$, $n = 0,1,\ldots$. 
The setup of Liu et al. can be thus restored 
when $\varphi$ is a difference operator with 
constant coefficients.  
\end{remark}

\section{Special solutions}

\subsection{Soliton solutions}

Soliton solutions provide a readily accessible playground 
in the theory of integrable hierarchies \cite{MJD-book}. 
It will be instructive to construct soliton solutions 
for the generalized ILW hierarchy as well.  To this end, 
we start from soliton solutions of the lattice KP hierarchy, 
and examine how they turn into solutions 
of the generalized ILW hierarchy.  

The soliton solutions of the lattice KP hierarchy 
can be obtained from those of the 2D Toda hierarchy 
\cite{UT84} by letting $\bar{\bst} = \bszero$. 
The tau function of the $M$-soliton solution 
thus takes the following form: 
\begin{align}
  \tau &= 1 + \sum_{i=1}^M a_ie_i 
         + \sum_{1\le i<j\le M}a_ia_jc_{ij}e_ie_j 
         + \sum_{1\le i<j<k\le M}a_ia_ja_kc_{ijk}e_1e_2e_3 
  \notag\\
       &\quad\mbox{} + \cdots 
         + a_1\cdots a_Mc_{1\cdots M}e_1\cdots e_M. 
  \label{lKP-soliton}
\end{align}
$a_i$'s are amplitude parameters and 
$e_i$'s are the following exponential factors: 
\[
  e_i = (p_i/q_i)^{s/\hbar}
        \exp\left(\hbar^{-1}(\xi(\bst,p_i) - \xi(\bst,q_i))\right). 
\]
$p_i$'s and $q_i$'s are another set of parameters 
with mutually distinct values, 
and $c_{ij\cdots k}$'s are thereby defined as 
\[
  c_{ij} = \frac{(p_i-p_j)(q_i-q_j)}{(p_i-q_j)(q_i-p_j)},\quad 
  c_{i_1\cdots i_m} = \prod_{1\le a<b\le m}c_{i_ai_b}. 
\]

It is easy to find a set of conditions on the parameters 
under which this tau function satisfies 
the reduction condition (\ref{tau-red-cond}). 

\begin{prop}
If the parameters $p_i,q_i$ satisfy the equations 
\beq
  p_i^N - \nu\log p_i = q_i^N - \nu\log q_i,\quad 
  i = 1,\ldots,M,
  \label{pq-cond}
\eeq
the tau function (\ref{lKP-soliton}) satisfies 
the reduction condition (\ref{tau-red-cond}) 
to the generalized ILW hierarchy. 
\end{prop}

\begin{proof}
(\ref{pq-cond}) implies that 
\[
  p_i^N - q_i^N = \nu\log(p_i/q_i), 
\]
so that $e_i$'s can be reassembled as 
\[
\begin{aligned}
  e_i &= (p_i/q_i)^{s/\hbar}\exp\left(\hbar^{-1}(p_i^N - q_i^N)t_N\right)e_i'\\
      &= \exp\left(\hbar^{-1}(p_i^N - q_i^N)(t_N + s/\nu)\right)e_i', 
\end{aligned}
\]
where $e_i'$ is a factor that does not depend on $t_N$ and $s$. 
Thus the tau function (\ref{lKP-soliton}) depends 
on $t_N$ and $s$ through the linear combination $t_N + s/\nu$.  
\end{proof}

Note that avatars of the right hand side of (\ref{lam-z}) 
show up in (\ref{pq-cond}).  By letting $\nu = 0$, 
(\ref{pq-cond}) turns into the $N$-reduction condition 
\beq
  p_i^N = q_i^N 
  \label{nu=0-pq-cond}
\eeq
for soliton solutions of the lattice GD hierarchy.

\subsection{Dressing operators of Okounkov-Pandharipande}

A special solution of quite different nature is hidden 
in the work of Okounkov and Pandharipande \cite{OP02} 
on the equivariant Gromov-Witten theory of $\CC\PP^1$.  
They use what they call ``dressing operators'' 
to convert a fermionic formula of the generating function 
of equivariant Gromov-Witten invariants 
to the standard fermionic formalism 
of 2D Toda tau functions \cite{TT95,Takasaki18}. 
In our previous work \cite{Takasaki21}, those operators 
(denoted by $V$ and $\bar{V}$ therein) are reformulated 
as difference operators in the spatial variable $s$, 
and used to explain how the equivariant Toda hierarchy 
emerges in the equivariant Gromov-Witten theory 
of $\CC\PP^1$.  

$V$ is a difference operator of the form 
\[
  V = 1 + \sum_{n=1}^\infty v_n\Lambda^{-n}
\]
and constructed to satisfy the intertwining relation 
\beq
  (\Lambda^N + H - \nu\log\Lambda)V 
  = V(\Lambda^N - \nu\log\Lambda), 
  \label{V-rel}
\eeq
where $H$ is the multiplication operator 
\footnote{The definition of $H$ is slightly different 
from our previous work \cite{Takasaki21} in two aspects. 
First, $s$ is rescaled by $\hbar$ so as to preserve 
the commutation relation $[\log\Lambda,s] = 1$. 
Second, the constant term is modified from $-1/2$ to $1/2$ 
to be consistent with the relation between the wave function 
and the tau function in this paper.}
\[
  H = s/\hbar + 1/2 
\]
that corresponds to the energy operator 
in the fermionic formalism.  More precisely, 
the work of Okounkov and Pandharipande \cite{OP02} 
amounts to the case of $N = 1$; the case of $N > 1$ 
is an orbifold generalization \cite{Johnson09}.

Let us briefly recall the construction of $V$ 
\cite{Takasaki21}.  $V$ is constructed order-by-order 
in the $\nu$-expansion 
\[
  V = \sum_{k=0}^\infty \nu^kV_k. 
\]
By this expansion, the intertwining relation (\ref{V-rel}) 
can be decomposed into an infinite number of equations 
of the following form: 
\begin{align}
  (\Lambda^N + H)V_0, &= V_0\Lambda^N \label{V-eq0}\\
  (\Lambda^N + H)V_k - \log\Lambda V_{k-1} 
    &= V_k\Lambda^N - V_{k-1}\log\Lambda, 
    \quad k = 1,2,\ldots. \label{V-eqk}
\end{align}
We can multiply both sides of (\ref{V-eqk}) by $V_0^{-1}$ 
and use the first equation (\ref{V-eq0}) 
to rewrite (\ref{V-eqk}) as 
\beq
  [\Lambda^N, V_0^{-1}V_k] = V_0^{-1}[\log\Lambda, V_{k-1}], 
  \label{V-eqk2}
\eeq
which is simpler than (\ref{V-eqk}). 
Starting with (\ref{V-eq0}), as explained below, 
we can solve these equations step-by-step 
and find that the $k$-th operator $V_k$ 
can be expressed as 
\[
  V_k = \sum_{n=kN}^\infty v_{kn}\Lambda^{-n} 
\]
with the coefficients $v_{kn}$ being polynomials in $s$. 
Consequently, $V$ becomes a difference operator of the form 
\[
  V = 1 + \sum_{n=1}^\infty \sum_{k=0}^{[n/N]} v_{kn}\Lambda^{-n}, 
\]
where $[n/N]$ denotes the integral part of $n/N$. 
Thus the coefficients of $V$ itself turn out to be 
polynomials in $s$.  

The operator equations (\ref{V-eq0}) and (\ref{V-eqk2}) 
consist of an infinite number of difference equations 
for the coefficients $v_{kn}$.  To find a solution 
of this system of equations, we have to solve 
a difference equation of the form 
\beq
  v(s+N\hbar) - v(s) = f(s), 
  \label{diffv-eq}
\eeq
for $v(s)$ repeatedly, where $f(s)$ is a given polynomial. 
As far as the right hand side is a polynomial in $s$, 
we can use the difference identities  
\[
  B_k(x+1) - B_k(x) = kx^{k-1}
\]
of the Bernoulli polynomials $B_k(x)$, 
defined by the generating function 
\[
  \frac{te^{xt}}{e^t - 1} = \sum_{k=0}^\infty B_k(x)\frac{t^k}{k!}, 
\]
to find a polynomial solution $v(s)$ of (\ref{diffv-eq}). 
\footnote{Our previous work \cite{Takasaki21} 
uses a different set of polynomials.} 

The intertwining relation readily implies that 
the operator 
\beq
  U = e^{\xi(\bst,\Lambda)/\hbar}V^{-1} 
  \label{U-V}
\eeq
satisfies the reduction condition (\ref{U-red-cond}) with 
\[
  \varphi = \Lambda^N + H. 
\]
The factorization problem (\ref{fac-prob})  thereby 
generates a solution of the generalized ILW hierarchy. 
\footnote{Solving the factorization problem explicitly 
is a difficult task.  The solution is encoded 
in the fermionic formula of tau functions 
\cite{TT95,Takasaki18}. 
}

\begin{remark}
A more interesting choice of the $U$-operator will be 
\beq
  U = e^{\xi(\bst,\Lambda)/\hbar}V^{-1}e^{\Lambda^N/N}, 
  \label{U-eqGWleft}
\eeq
which is the left half of the $U$-operator (\ref{U-eqGW}) 
for the equivariant Gromov-Witten theory of $\CC\PP^1$. 
This operator satisfies the reduction condition 
(\ref{U-red-cond}) with 
\[
  \varphi = H. 
\]
Actually, (\ref{U-V}) and (\ref{U-eqGWleft}) yield 
the same factor $W$ in the factorization problem 
(\ref{fac-prob}).  The right action $U \to UC$ 
of the $U$-operator by a difference operator of the form 
\[
  C = \sum_{n=0}^\infty c_n\Lambda^n,\quad 
  c_0 \not= 0,\quad 
  \frac{\rd C}{\rd t_k} = 0 \quad, k = 1,2,\ldots
\]
leaves $W$ invariant and changes $\bar{W}$ 
as $\bar{W} \to \bar{W}C$.  
\end{remark}

\subsection{Generalization}

The foregoing construction of $V$ (and the special solution 
of the generalized ILW hierarchy) can be extended 
to the case where $\varphi$ takes a form more general 
than $\varphi = \Lambda^N + H$.  Let us assume 
that $\varphi$ is a difference operator of the form 
\beq
  \varphi = \Lambda^N + \sum_{n=1}^N\varphi_n\Lambda^{N-n}  
  \label{varphi-gen}
\eeq
and that $\varphi_n$'s do not depend on $\nu$.  
The second assumption can be relaxed, but we impose it 
to simplify the subsequent computations.  
The problem is to construct a difference operator 
\[
  V = 1 + \sum_{n=1}^\infty v_n\Lambda^{-n} 
\]
that satisfies the intertwining relation 
\beq
  (\varphi- \nu\log\Lambda)V = V(\Lambda^N - \nu\log\Lambda). 
  \label{V-rel-gen}
\eeq

Plugging the $\nu$-expansion 
\[
  V = \sum_{k=0}^\infty \nu^kV_k, 
\]
into (\ref{V-rel-gen}) yields the following generalization 
of (\ref{V-eq0}) and (\ref{V-eqk}): 
\begin{align}
  \varphi V_0, &= V_0\Lambda^N \label{V-eq0-gen}\\
  \varphi V_k - \log\Lambda V_{k-1} 
    &= V_k\Lambda^N - V_{k-1}\log\Lambda, 
    \quad k = 1,2,\ldots. \label{V-eqk-gen}
\end{align}
The second equations can be converted to the equations 
\[
  [\Lambda^N, V_0^{-1}V_k] = V_0^{-1}[\log\Lambda, V_{k-1}] 
\]
of the same form as (\ref{V-eqk2}).  
If $\varphi_n$'s are polynomials in $s$, 
we can solve these equations in much the same way 
as the foregoing case.  Thus we arrive at the following result. 

\begin{prop}
If $\varphi$ takes the form as shown in (\ref{varphi-gen}) 
and $\varphi_n$'s are polynomials in $s$, 
there is a difference operator $V$ of the form 
\[
  V = 1 + \sum_{n=1}^\infty\sum_{k=0}^{[n/N]}v_{kn}\Lambda^{-n}
\]
such that the intertwining relation (\ref{V-rel-gen}) holds 
and $v_{kn}$'s are polynomials in $s$.  The operator 
\beq
  U = e^{\xi(\bst,\Lambda)/\hbar}V^{-1}
\eeq
satisfies the reduction condition (\ref{U-red-cond}) 
to the generalized ILW hierarchy. 
\end{prop}

\begin{remark}
We can also construct $U_{\rm{in}} = V^{-1}$ directly 
by the $\nu$-expansion 
\[
  U_{\rm{in}} = \sum_{k=0}^\infty \nu^kU^{\rm{in}}_k. 
\]
$U^{\rm{in}}_k$'s are required to satisfy the equations 
\[
\begin{aligned}
  \Lambda^NU^{\rm{in}}_0 &= U^{\rm{in}}_0\varphi, \\
  \Lambda^NU^{\rm{in}}_k - \log\Lambda U^{\rm{in}}_{k-1}   
    &= U^{\rm{in}}_k\varphi - U^{\rm{in}}_{k-1}\log\Lambda, 
    \quad k = 1,2,\ldots. 
\end{aligned}
\]
Solving these equations for $U^{\rm{in}}_k$'s 
is mostly parallel to the construction of $V_k$'s.  
\end{remark}

\begin{remark}
It is an interesting issue to construct a solution 
of the bigraded equivariant Toda hierarchy 
from the foregoing generalization of $V$ 
and its partner $\bar{V}$.  $\bar{V}$ is required 
to satisfy the intertwining relation 
\beq
  (\Lambda^{-\bar{N}} - \nu\log\Lambda)\bar{V} 
  = \bar{V}(\bar{\varphi} - \nu\log\Lambda), 
\eeq
where $\bar{N}$ is another positive integer, 
and $\bar{\varphi}$ is a difference operator 
of the form 
\[
  \bar{\varphi} = \Lambda^{-\bar{N}} 
    + \sum_{n=1}^{\bar{N}}\bar{\varphi}_n\Lambda^{n-\bar{N}} 
\]
with polynomial coefficients $\bar{\varphi}_n$.  
If $\varphi$ and $\bar{\varphi}$ are intertwined 
by another difference operator $S$ as 
\beq
  (\varphi - \nu\log\Lambda)S 
  = S(\bar{\varphi} - \nu\log\Lambda), 
  \label{S-rel}
\eeq
one can build the $U$-operator 
\beq
  U = e^{\xi(\bst,\Lambda)/\hbar}V^{-1}S 
      \bar{V}^{-1}e^{-\xi(\bar{\bst},\Lambda^{-1})/\hbar} 
\eeq
that satisfies the reduction condition (\ref{eq-U-cond}) 
from the 2D Toda hierarchy to the equivariant 
bigraded Toda hierarchy.  The factorization problem 
for the 2D Toda hierarchy thereby generates a solution 
of the equivariant bigraded Toda hierarchy 
of type $(N,\bar{N})$.  In order to accommodate 
the equivariant Gromov-Witten theory of $\CC\PP^1$, 
however, one has to modify the intertwining relation 
(\ref{S-rel}) by a constant term \cite{MT07,Takasaki21}
(see Appendix, Remark \ref{rem:UQ}). 
\end{remark}

\section{Limit to extended lattice GD hierarchy}

Let us turn to the issue of the limit 
to the lattice GD hierarchy.  
To avoid a trivial situation, we here consider 
the case where $N > 1$.  

In the naive limit as $\nu \to 0$, 
the reduction condition (\ref{red-cond}) becomes 
the $N$-reduction condition 
\beq
  (L^N)_{<0} = 0 
  \label{N-red-cond}
\eeq
to the lattice GD hierarchy.  Since this condition 
implies that $B_{kN} = L^{kN}$, $k = 1,2,\ldots$, 
time evolutions with respect to $t_{kN}$'s 
are trivialized, namely, 
\[
  \frac{\rd L}{\rd t_{kN}} = 0, \quad 
  \frac{\rd W}{\rd t_{kN}} = 0. 
\]
More precisely, the naive limit means that 
all dynamical variables, i.e., the coefficients 
$u_n = u_n(\hbar,\nu,s,\bst)$ and $w_n = w_n(\hbar,\nu,s,\bst)$ 
of the Lax and the dressing operators have a smooth limit 
\beq
  u_n = u_n^{(0)} + O(\nu), \quad w_n = w_n^{(0)} + O(\nu) 
  \label{naive-limit}
\eeq
as $\nu \to 0$, where $u_n^{(0)} = u^{(0)}(\hbar,s,\bst)$ 
and $w_n^{(0)} = w_n^{(0)}(\hbar,s,\bst)$ are functions 
independent of $\nu$.  
Although the same notations are used, $L$ and $W$ 
in the equations shown above denote the operators 
with $u_n$ and $w_n$ being replaced by $u_n^{(0)}$ and $w_n^{(0)}$. 

We argue in the following that a more careful prescription 
of the $\nu \to 0$ limit leads to emergence 
of an extended set of flows in place of 
the trivialized $t_{kN}$-flows.  
The outcome is the extended lattice GD hierarchy introduced 
in our previous work \cite{Takasaki22}.

\subsection{Extended lattice GD hierarchy}

Before formulating the scaling limit, let us recall 
the construction of the extended lattice GD hierarchy.  

Let $\frakL$ denote the reduced Lax operator 
\[
  \frakL = L^N = B_N 
  = \Lambda^N + b_1\Lambda^{N-1} + \cdots  + b_N
\]
under the reduction condition (\ref{N-red-cond}). 
$\frakL$ satisfies the reduced Lax equations 
\[
  \hbar\frac{\rd\frakL}{\rd t_k} = [B_k,\frakL]. 
\]
The lattice KP hierarchy is thus reduced 
to evolution equations for $b_n$'s.  

The extended lattice GD hierarchy is obtained 
by adding an infinite number of time variables 
$\bsx = (x_k)_{k=1}^\infty$ and Lax equations 
\beq
  \hbar\frac{\rd\frakL}{\rd x_k} = [C_k,\frakL]
  \label{x-Laxeq}
\eeq
in place of the trivialized $t_{kN}$-flows.  
The generators $C_k$ of these flows are given by 
\beq
  C_k = \left(L^{kN}\log L\right)_{\geq 0}. 
  \label{Ck}
\eeq
Since 
\[
  L^{kN}\log L = W\Lambda^{kN}\log\Lambda W^{-1} 
  = \frakL^k\log\Lambda - \frakL^k\hbar\frac{\rd W}{\rd s}W^{-1}, 
\]
we can rewrite $C_k$ into the expression 
\beq
  C_k = \frakL^k\log\Lambda 
      - \left(\frakL^k\hbar\frac{\rd W}{\rd s}W^{-1}\right)_{\ge 0} 
  \label{Ck-bis}
\eeq
presented in our previous work \cite{Takasaki22}.  
Because of the presence of the logarithmic terms, 
these flows are called logarithmic flows.  
Since $C_0 = \log\Lambda$, the $x_0$-flow can be identified  
with the translation in the $s$-direction.  

The Sato equations,too, are extended to the logarithmic flows as 
\beq
  \hbar\frac{\rd W}{\rd x_k} = C_kW - W\Lambda^{kN}\log\Lambda. 
  \label{x-Satoeq}
\eeq
These equations imply that $\frakL = W\Lambda^NW^{-1}$ 
satisfies the Lax equations (\ref{x-Laxeq}).

\subsection{Scaling limit of Lax and Sato equations} 

We consider the following scaling transformations 
of time variables from $\{t_k\}_{k=1}^\infty$ 
to $\{T_k\}_{k \not\equiv 0 \mod N}$ and $\{X_k\}_{k=1}^\infty$: 
\beq
  t_k = \begin{cases}
        T_k        &\text{if $k \not\equiv 0 \mod N$},\\
        X_{k/N}/\nu &\text{if $k \equiv 0 \mod N$}.
        \end{cases} 
  \label{t-TX}
\eeq
The derivatives are thus transformed as 
\begin{align}
  \frac{\rd}{\rd t_k} &= \frac{\rd}{\rd T_k} 
  \quad\text{for $k \not\equiv 0\mod N$}, 
  \label{t-T}\\
  \frac{\rd}{\rd t_{kN}} &= \nu\frac{\rd}{\rd X_k}
  \quad\text{for $k = 1,2,\ldots$}. 
  \label{t-X}
\end{align}
We assume that $u_n$ and $w_n$, viewed as functions 
of the rescaled variables $T_k,X_k$, have a smooth limit 
\beq
\begin{aligned}
  u_n &= u_n^{(0)}(\hbar,s,T_1,\ldots,T_{N-1},X_1,
         T_{N+1},\ldots,T_{2N-1},X_2,\ldots) + O(\nu),\\
  w_n &= w_n^{(0)}(\hbar,s,T_1,\ldots,T_{N-1},X_1,
         T_{N+1},\ldots,T_{2N-1},X_2,\ldots) + O(\nu) 
  \label{refined-limit}
\end{aligned}
\eeq
as $\nu \to 0$. This condition is stronger than 
the condition (\ref{naive-limit}) for the naive limit. 

Let $L(0)$ and $W(0)$ denote the difference operators 
with $u_n$ and $w_n$ being replaced by $u_n^{(0)}$ and $w_n^{(0)}$: 
\[
  L(0) = \Lambda + \sum_{n=1}^\infty u_n^{(0)}\Lambda^{1-n},\quad 
  W(0) = 1 + \sum_{n=1}^\infty w_n^{(0)}\Lambda^{-n}. 
\]
$L(0)$ and $W(0)$ are connected by the dressing relation
\[
  L(0) = W(0)\Lambda W(0)^{-1}. 
\]
The reduction condition (\ref{red-cond}) for $L$ 
turns into the condition 
\beq
  \left(L(0)^N\right)_{<0} = 0 
  \label{L0-red-cond}
\eeq
for $L(0)$.  This is again a reduction condition 
to the lattice GD hierarchy.  Moreover, we can derive 
a set of Lax and Sato equations for $L(0)$ and $W(0)$ 
as follows. 

\begin{prop}
$W(0)$ satisfies the Sato equations 
\begin{align}
  \hbar\frac{\rd W(0)}{\rd T_k} &= B_k(0)W(0) - W(0)\Lambda^k 
  \quad \text{for $k \not\equiv 0 \mod N$}, 
  \label{T-Satoeq}\\
  \hbar\frac{\rd W(0)}{\rd X_k}  
    &= kC_k(0)W(0) - kW(0)\Lambda^{kN}\log\Lambda 
  \quad \text{for $k = 1,2,\ldots$},  
  \label{X-Satoeq}
\end{align}
where 
\[
  B_k(0) = \left(L(0)^k\right)_{\ge 0},\quad 
  C_k(0) = \left(L(0)^{(k-1)N}\log L(0)\right)_{\ge 0}. 
\]
\end{prop}

\begin{proof}
In view of (\ref{t-T}), the Sato equation (\ref{T-Satoeq}) 
for $W(0)$ is an immediate consequence 
of the Sato equations (\ref{Satoeq}) for $W$ 
with respect to $t_k$, $k \not\equiv 0 \mod N$.  
In order to derive (\ref{X-Satoeq}), 
we consider the operator 
\[
  \calL_k = (L^N - \nu\log L)^k 
  = L^{kN} - k\nu L^{(k-1)N}\log L + O(\nu^2). 
\]
The reduction condition (\ref{red-cond}) implies that 
\[
  \calL_k = \left((L^N - \nu\log L)^k\right)_{\ge 0} 
  = B_{kN} - k\nu C_k + O(\nu^2),
\]
where 
\[
  C_k = \left(L^{(k-1)N}\log L\right)_{\ge 0}. 
\]
Let us examine the operator identity 
\beq
  \calL_kW = W(\Lambda^N - \nu\log\Lambda)^k. 
  \label{calLW=W...}
\eeq
We can use the Sato equation (\ref{Satoeq}) 
and the scaling relation (\ref{t-X}) to rewrite 
the left hand side of this identity as 
\[
\begin{aligned}
  \calL_kW 
  &= B_{kN}W - k\nu C_kW + O(\nu^2) \\
  &= \hbar\frac{\rd W}{\rd t_{kN}} + W\Lambda^{kN} - k\nu C_kW + O(\nu^2)\\
  &= \hbar\nu\frac{\rd W}{\rd X_k} + W\Lambda^{kN} - k\nu C_kW + O(\nu^2). 
\end{aligned}
\]
The right hand side of (\ref{calLW=W...}) 
can be expanded as 
\[
  W(\Lambda^N - \nu\log\Lambda)^k 
  = W\Lambda^{kN} - k\nu W\Lambda^{(k-1)N}\log\Lambda + O(\nu^2). 
\]
We can thus extract the coefficients of $\nu$ 
in the $\nu$-expansion of (\ref{calLW=W...}) 
to obtain (\ref{X-Satoeq}). 
\end{proof}

\begin{cor}
$L(0)$ satisfies the Lax equations 
\begin{align}
  \hbar\frac{\rd L(0)}{\rd T_k} &= [B_k(0),L(0)] 
  \quad \text{for $k \not\equiv 0 \mod N$}, 
  \label{T-Laxeq}\\
  \hbar\frac{\rd L(0)}{\rd X_k} &= [kC_k(0),L(0)] 
  \quad \text{for $k = 1,2,\ldots$}. 
  \label{X-Laxeq}
\end{align}
\end{cor}

We thus obtain two sets of flows defined 
by the Lax equations (\ref{T-Laxeq}), (\ref{X-Laxeq}) 
and the Sato equations (\ref{T-Satoeq}), (\ref{X-Satoeq}) 
alongside the reduction condition (\ref{L0-red-cond}). 
These equations are essentially the same as those 
of our previous work \cite{Takasaki22}, 
cf. (\ref{x-Laxeq}) -- (\ref{x-Satoeq}), 
except that the time variables of the logarithmic flows 
are rescaled and renumbered as 
\[
  x_{k-1} = kX_k, \quad k = 1,2,\ldots. 
\]

\subsection{Scaling limit of soliton solutions}

We can apply the foregoing prescription 
of scaling limit to the soliton solution 
(\ref{lKP-soliton}) as well. 
Suppose that the parameters $p_i,q_i$ 
of the soliton solution (\ref{lKP-soliton}) 
depend on $\nu$, behave as 
\[
  p_i = p_i(0) + O(\nu), \quad 
  q_i = q_i(0) + O(\nu)
\]
as $\nu \to 0$, and satisfy the reduction condition 
(\ref{pq-cond}).  $p_i(0)$ and $q_i(0)$ thereby 
satisfies the reduction condition 
\[
  p_i(0)^N = q_i(0)^N
\]
to the lattice GD hierarchy.  

In this setup, we do the scaling transformation (\ref{t-TX}) 
of time variables.  The exponential factors $e_i$ 
in the tau function turn out to have the following limit 
as $\nu \to 0$. 

\begin{prop}
\begin{align}
  \lim_{\nu\to 0}e_i 
  &= \left(\frac{p_i(0)}{q_i(0)}\right)^{s/\hbar}
     \exp\left(\hbar^{-1}\sum_k{}'\,
     \left(p_i(0)^k - q_i(0)^k\right)T_k\right) \notag\\
  &\mbox{}\times \exp\left(\hbar^{-1}\sum_{k=1}^\infty
     k\left(p_i(0)^{(k-1)N}\log p_i(0) - q_i(0)^{(k-1)N}\log q_i(0)\right)
     X_k\right), 
  \label{ei-limit}
\end{align}
where $\sum_k'$ denotes the sum over $k = 1,2,\dots$, 
$k \not\equiv 0 \mod N$. 
\end{prop}

\begin{proof}
Let us consider the obvious consequence 
\[
  (p_i^N - \nu\log p_i)^k = (q_i^N - \nu\log q_i)^k
\]
of the reduction condition (\ref{pq-cond}).  
Both sides can be expanded as 
\[
\begin{aligned}
  (p_i^N - \nu\log p_i)^k 
  &= p_i^{kN} - k\nu p_i^{(k-1)N}\log p_i + O(\nu^2),\\
  (q_i^N - \nu\log q_i)^k 
  &= q_i^{kN} - k\nu q_i^{(k-1)N}\log q_i + O(\nu^2),
\end{aligned}
\]
hence 
\[
\begin{aligned}
  p_i^{kN} - q_i^{kN} 
  &= \nu k(p_i^{(k-1)N}\log p_i - q_i^{(k-1)N}\log q_i) + O(\nu^2) \\
  &= \nu k(p_i(0)^{(k-1)N}\log p_i(0) - q_i(0)^{(k-1)N}\log q_i(0)) 
  + O(\nu^2). 
\end{aligned}
\]
Consequently, the part of $\xi(\bst,p_i) - \xi(\bst,q_i)$ 
containing $t_{kN}$'s becomes 
\[
  \sum_{k=1}^\infty (p_i^{kN} - q_i^{kN})t_{kN} 
  = \sum_{k=1}^\infty k(p_i(0)^{(k-1)N}\log p_i(0) 
        - q_i(0)^{(k-1)N}\log q_i(0))X_k 
     + O(\nu). 
\]
The other part of $\xi(\bst,p_i) - \xi(\bst,q_i)$ 
and $(p_i/q_i)^{s/\hbar}$ have the simpler asymptotic form 
\[
\begin{aligned}
  (p_i/q_i)^{s/\hbar} &= (p_i(0)/q_i(0))^{s/\hbar} + O(\nu),\\
  \sum_k{}'(p_i^k - q_i^k)t_k 
  &= \sum_k{}'(p_i(0)^k - q_i(0)^k)T_k + O(\nu). 
\end{aligned}
\]
\end{proof}

\begin{cor}
In the limit as $\nu \to 0$, the tau function 
(\ref{lKP-soliton}) converges to 
\begin{align}
  \calT &= 1 + \sum_{i=1}^M a_iE_i 
         + \sum_{1\le i<j\le M}a_ia_jC_{ij}E_iE_j 
         + \sum_{1\le i<j<k\le M}a_ia_ja_kC_{ijk}E_1E_2E_3 
  \notag\\
       &\quad\mbox{} + \cdots 
         + a_1\cdots a_MC_{1\cdots M}E_1\cdots E_M, 
  \label{xlGD-soliton}
\end{align}
where 
\[
  C_{ij} = \frac{(p_i(0)-p_j(0))(q_i(0)-q_j(0))}
          {(p_i(0)-q_j(0))(q_i(0)-p_j(0))},\quad 
  C_{i_1\cdots i_m} = \prod_{1\le a<b\le m}C_{i_ai_b}, 
\]
and $E_i$ denotes the right hand side of (\ref{ei-limit}). 
\end{cor}

Thus, as anticipated, we obtain the tau function 
(\ref{xlGD-soliton}) of the $N$-soliton solution 
of the extended lattice GD hierarchy.

\section{Conclusion}

We have formulated the ILW hierarchy and its generalization 
as reductions of the lattice KP hierarchy.  
The integrability of the lattice KP hierarchy is inherited 
by these reduced systems.  In particular, all solutions 
can be described by the factorization problem 
for the lattice KP hierarchy. The $U$-operator therein 
is required to satisfy a reduction condition.  
The situation is thus parallel to the equivariant 
1D/bigraded Toda hierarchy. 

We have uncovered some features hidden 
in these somewhat exotic integrable hierarchies.  

First, we can thereby explain an origin 
of the logarithmic flows of the lattice GD hierarchy 
\cite{Takasaki22} and the 1D/bigraded Toda hierarchy 
\cite{CDZ04,Carlet06}.  These integrable hierarchies 
are a naive limit of the generalized ILW hierarchy 
and the equivariant Toda hierarchy 
as the parameter $\nu$ tends to $0$. In this limit, 
part of the flows are trivialized, and the logarithmic flows 
are added \textit{by hand} in place of these trivialized flows.  
We have found that the logarithmic flows can be \textit{derived} 
by a more refined formulation of the limit. 

Second, a special solutions related to the equivariant 
Toda hierarchy can be obtained from the factorization problem.  
The $U$-operator therein is built from one of 
the dressing operators of Okounkov and Pandharipande 
\cite{OP02,Johnson09,Takasaki21}.  The dressing operators 
were originally introduced for the construction 
of a solution of the equivariant Toda hierarchy.  
This indicates a somewhat puzzling, but quite intriguing 
relation between the generalized ILW hierarchy 
and the equivariant Toda hierarchy.  Moreover, 
we have shown a generalization of the dressing operators, 
which can lead to new special solution of these 
integrable hierarchies.

\subsection*{Acknowledgements}

This work is partly supported by the JSPS Kakenhi Grant 
JP18K03350 and JP21K03261. 

\appendix

\section{Equivariant Toda hierarchy}

\subsection{Reduction from 2D Toda hierarchy}

The 2D Toda hierarchy consists of the Lax equations 
\[
\begin{gathered}
  \hbar\frac{\rd L}{\rd t_k} = [B_k,L],\quad 
  \hbar\frac{\rd L}{\rd\bar{t}_k} = [\bar{B}_k,L],\\
  \hbar\frac{\rd \bar{L}}{\rd t_k} = [B_k,\bar{L}],\quad 
  \hbar\frac{\rd \bar{L}}{\rd\bar{t}_k} = [\bar{B}_k,\bar{L}],
\end{gathered}
\]
for two Lax operators $L$ and $\bar{L}$ of the form 
mentioned in the main text.  $B_k$ and $\bar{B}_k$ 
are defined as 
\[
  B_k = (L^k)_{\geq 0},\quad \bar{B}_k = (\bar{L}^{-k})_{<0}. 
\]

The Lax operators can be expressed in the dressed form 
\[
    L = W\Lambda W^{-1},\quad 
  \bar{L} = \bar{W}\Lambda\bar{W}^{-1}. 
\]
These Lax equations are thereby converted 
to the Sato equations 
\[
\begin{gathered}
  \frac{\rd W}{\rd t_k} = B_kW - W\Lambda^k,\quad 
  \frac{\rd W}{\rd\bar{t}_k} = \bar{B}_kW,\\
  \frac{\rd\bar{W}}{\rd t_k} = B_k\bar{W},\quad 
  \frac{\rd\bar{W}}{\rd\bar{t}_k} = \bar{B}_k\bar{W} - W\Lambda^{-k}. 
\end{gathered}
\]
The dressing operators are also characterized 
by the factorization problem 
\[
  U = W^{-1}\bar{W}, 
\]
where $U$ satisfies the evolution equations 
\[
  \hbar\frac{\rd U}{\rd t_k} = \Lambda^kU,\quad 
  \hbar\frac{\rd U}{\rd\bar{t}_k} = - U\Lambda^{-k}. 
\]
Given the initial values 
$W_{\rm{in}} = W|_{\bst=\bar{\bst}=\bszero}$ and 
$\bar{W}_{\rm{in}} = \bar{W}|_{\bst=\bar{\bst}=\bszero}$, 
the initial value problem of the Sato equations 
can be solved by the factorization problem with 
\beq
  U = e^{\xi(\bst,\Lambda)/\hbar}W_{\rm{in}}^{-1}\bar{W}_{\rm{in}}
      e^{-\xi(\bar{\bst},\Lambda^{-1})/\hbar}. 
\eeq

The equivariant bigraded Toda hierarchy 
of type $(N,\bar{N})$ \cite{MT07} 
can be derived by the reduction condition 
\beq
  L^N - \nu\log L = \bar{L}^{-\bar{N}} - \nu\log\bar{L}. 
  \label{eq-red-cond}
\eeq
The original equivariant Toda hierarchy amounts 
to the $N = \bar{N} = 1$ case \cite{Getzler04}.  
Since 
\[
\begin{aligned}
  L^N - \nu\log L 
  &= B_N - \nu\log\Lambda 
     + \text{negative powers of $\Lambda$},\\
  \bar{L}^{-\bar{N}} - \nu\log\bar{L}
  &= \bar{B}_N - \nu\log\Lambda 
     + \text{non-negative powers of $\Lambda$},
\end{aligned}
\]
the reduction condition yields the reduced Lax operator 
\beq
  \frakL = B_N + \bar{B}_{\bar{N}} - \nu\log\Lambda 
\eeq
that satisfies the reduced Lax equations 
\beq
  \hbar\frac{\rd\frakL}{\rd t_k} = [B_k,\frakL],\quad 
  \hbar\frac{\rd\frakL}{\rd\bar{t}_k} = [\bar{B}_k,\frakL]. 
\eeq
In the language of the factorization problem, 
the reduction condition reads 
\beq
  (\Lambda^N - \nu\log\Lambda)U 
  = U(\Lambda^{-\bar{N}} - \nu\log\Lambda). 
  \label{eq-U-cond}
\eeq

The reduction condition (\ref{eq-red-cond}) can be 
translated to the dressing operators 
and the tau function as well. In particular, 
the reduction condition for the tau function reads 
\beq
  \frac{\rd\tau}{\rd t_N} + \frac{\rd\tau}{\rd\bar{t}_{\bar{N}}} 
    - \nu\frac{\rd\tau}{\rd s} = 0, 
  \label{eq-tau-cond}
\eeq
hence the tau function depends on $t_N$, $\bar{t}_{\bar{N}}$ 
and $s$ through the linear combinations 
$t_N + s/\nu$ and $\bar{t}_{\bar{N}} + s/\nu$: 
\beq
  \tau = f(\hbar,\nu,t_1,\ldots,t_{N-1},t_N + s/\nu,t_{N+1},\ldots, 
         \bar{t}_1,\ldots,\bar{t}_{\bar{N}-1},
         \bar{t}_{\bar{N}} + s/\nu,\bar{t}_{\bar{N}+1},\ldots). 
  \label{eq-red-tau}
\eeq

\begin{remark}
\label{rem:UQ}
In order to deal with the equivariant Gromov-Witten theory 
of $\CC\PP^1$, one has to modify (\ref{eq-red-cond}) 
and (\ref{eq-U-cond}) as 
\begin{align}
  L^N - \nu\log L 
    &= \bar{L}^{-\bar{N}} - \nu\log\bar{L} - \nu\log Q, \\
  (\Lambda^N - \nu\log\Lambda)U 
    &= U(\Lambda^{-\bar{N}} - \nu\log\Lambda - \nu\log Q),  
\end{align}
where $Q$ is a constant \cite{MT07,Takasaki22}.  
The reduced form (\ref{eq-red-tau}) 
of the tau function is accordingly modified as 
\beq
  \tau 
  = Q^{(s/\hbar)^2/2}f(\hbar,\nu,t_1,\ldots,t_{N-1},t_N + s/\nu,t_{N+1},
    \ldots,\bar{t}_1,\ldots,\bar{t}_{\bar{N}-1},
    \bar{t}_{\bar{N}} + s/\nu,\bar{t}_{\bar{N}+1},\ldots). 
\eeq
The solution of the equivariant Gromov-Witten theory 
is obtained from the $U$-operator 
\beq
  U = e^{\xi(\bst,\Lambda)/\hbar}V^{-1}e^{\Lambda^N/N}Q^H 
      e^{\Lambda^{-\bar{N}}/\bar{N}}\bar{V}^{-1}
      e^{-\xi(\bar{\bst},\Lambda^{-1})/\hbar}, 
  \label{U-eqGW}
\eeq
where $\bar{V}$ is a partner of $V$ as explained in Section 4, 
and satisfies the intertwining relation 
\beq
  (\Lambda^{-\bar{N}} - \nu\log\Lambda)\bar{V}
  = \bar{V}(\Lambda^{-\bar{N}} + H - \nu\log\Lambda). 
  \label{Vbar-rel}
\eeq
The operator 
\beq
  S = e^{\Lambda^N/N}Q^He^{\Lambda^{-\bar{N}}/\bar{N}}
\eeq
intertwines $\Lambda^N + H - \nu\log\Lambda$ 
on the left hand side of (\ref{V-rel}) and 
$\Lambda^{-\bar{N}} + H - \nu\log\Lambda$ 
on the right side of (\ref{Vbar-rel}) 
up to a constant term as 
\beq
  (\Lambda^N + H - \nu\log\Lambda)S 
  = S(\Lambda^{-\bar{N}} + H - \nu\log\Lambda - \nu\log Q). 
\eeq
\end{remark}

\subsection{Limit to extended bigraded Toda hierarchy} 

The naive limit of (\ref{eq-red-cond}) as $\nu \to 0$ 
is the reduction condition 
\[
  L^N = \bar{L}^{-\bar{N}}
\]
to the bigraded Toda hierarchy of type $(N,\bar{N})$, 
which becomes the 1D Toda hierarchy when $N = \bar{N} = 1$. 
In this reduced hierarchy, the sum of the $t_{kN}$-flow 
and the $\bar{t}_{k\bar{N}}$-flow is trivialized 
for $k = 1,2,\ldots$, namely, 
\beq
  \frac{\rd L}{\rd t_{kN}} 
    + \frac{\rd L}{\rd\bar{t}_{k\bar{N}}} = 0,\quad
  \frac{\rd\bar{L}}{\rd t_{kN}} 
    + \frac{\rd\bar{L}}{\rd\bar{t}_{k\bar{N}}} = 0
\eeq
and 
\beq
  \frac{\rd W}{\rd t_{kN}} 
    + \frac{\rd W}{\rd\bar{t}_{k\bar{N}}} = 0,\quad
  \frac{\rd\bar{W}}{\rd t_{kN}} 
    + \frac{\rd\bar{W}}{\rd\bar{t}_{k\bar{N}}} = 0. 
\eeq
The extended bigraded Toda hierarchy is constructed 
by adding logarithmic flows in place 
of these trivialized flows \cite{Carlet06,Li-etal09}. 

A more careful prescription of the $\nu \to 0$ limit 
enables us to derive the logarithmic flows. 
This prescription is based on the scaling 
transformations 
\begin{align}
  t_k &= \begin{cases}
    T_k &\text{if $k \not\equiv 0 \mod N$},\\
    T_k + X_{k/N}/\nu &\text{if $k \equiv 0 \mod N$},
    \end{cases} 
  \label{eq-t-TX}\\
  \bar{t}_k &= \begin{cases}
    \bar{T}_k &\text{if $k \not\equiv 0 \mod \bar{N}$},\\
    X_{k/\bar{N}}/\nu &\text{if $k \equiv 0 \mod \bar{N}$} 
    \end{cases}
  \label{eq-tbar-TbarX}
\end{align}
of the time variables from $\{t_k\}_{k=1}^\infty$ 
and $\{\bar{t}_k\}_{k=1}^\infty$ to $\{T_k\}_{k=1}^\infty$, 
$\{\bar{T}_k\}_{k\not\equiv 0\mod\bar{N}}$ 
and $\{X_k\}_{k=1}^\infty$. 
\footnote{The same transformations are used 
in the work of Okounkov and Pandharipande \cite{OP02}
to consider some particular elements 
in the equivariant cohomology of $\CC\PP^1$.}
The derivatives are accordingly transformed as 
\begin{align}
  \frac{\rd}{\rd t_k} &= \frac{\rd}{\rd T_k} 
  \quad\text{for $k = 1,2,\ldots$},\\
  \frac{\rd}{\rd\bar{t}_k} &= \frac{\rd}{\rd\bar{T}_k}  
  \quad\text{for $k \not\equiv 0 \mod \bar{N}$},\\
  \frac{\rd}{\rd\bar{t}_{k\bar{N}}} 
    &= \nu\frac{\rd}{\rd X_k} - \frac{\rd}{\rd T_{kN}}
  \quad\text{for $k = 1,2,\ldots$}. 
\end{align}
The dynamical variables $u_n$, etc., viewed 
to be functions of the new variables $T_k$, $\bar{T}_k$ 
and $X_k$, are assumed to have a smooth limit 
\[
  u_n = u^{(0)}(\hbar,s,\{T_k\},\{\bar{T}_k\},\{X_k\}) + O(\nu),
  \quad\text{etc.}
\]
as $\nu \to 0$.  

Let $L(0)$, $\bar{L}(0)$, $W(0)$ and $\bar{W}(0)$ 
denote the difference operators with 
the coefficients being replaced by $u_n^{(0)}$, etc.  
$L(0)$ and $\bar{L}(0)$ are expressed 
in the dressed form 
\[
  L(0) = W(0)\Lambda W(0)^{-1},\quad 
  \bar{L}(0) = \bar{W}\Lambda\bar{W}^{-1}
\]
and satisfies the reduction condition 
\beq
  L(0)^N = \bar{L}(0)^{-\bar{N}}
\eeq
to the bigraded Toda hierarchy of type $(N,\bar{N})$. 
These operators satisfy a set of Lax and Sato 
equations of the following form. 

\begin{prop}
$W(0)$ and $\bar{W}(0)$ satisfy the Sato equations 
\begin{align}
  \hbar\frac{\rd W(0)}{\rd T_k} &= B_k(0)W(0) - W(0)\Lambda^k 
    \quad \text{for $k = 1,2,\ldots$}, 
    \label{eq-TW-eq}\\
  \hbar\frac{\rd W(0)}{\rd\bar{T}_k} &= \bar{B}_k(0)W(0) 
    \quad \text{for $k \not\equiv 0 \mod \bar{N}$}, 
    \label{eq-TbarW-eq}\\
  \hbar\frac{\rd W(0)}{\rd X_k} 
      &= kC_k(0)W(0) - kW\Lambda^{(k-1)N}\log\Lambda 
    \quad \text{for $k = 1,2,\ldots$}, 
    \label{eq-XW-eq}
\end{align}
and
\begin{align}
  \hbar\frac{\rd\bar{W}(0)}{\rd T_k} &= B_k(0)\bar{W}(0) 
    \quad \text{for $k = 1,2,\ldots$}, 
    \label{eq-TWbar-eq}\\
  \hbar\frac{\rd\bar{W}(0)}{\rd\bar{T}_k} 
    &= \bar{B}_k(0)\bar{W}(0) - \bar{W}(0)\Lambda^{-k} 
    \quad \text{for $k \not\equiv 0 \mod \bar{N}$}, 
    \label{eq-TbarWbar-eq}\\
  \hbar\frac{\rd\bar{W}(0)}{\rd X_k} 
    &= kC_k(0)\bar{W}(0) - k\bar{W}\Lambda^{-(k-1)\bar{N}}\log\Lambda 
    \quad \text{for $k = 1,2,\ldots$}, 
    \label{eq-XWbar-eq}
\end{align}
where 
\[
\begin{gathered}
  B_k(0) = \left(L(0)^k\right)_{\ge 0},\quad 
  \bar{B}_k(0) = \left(\bar{L}(0)^{-k}\right)_{<0},\\
  C_k(0) = \left(L(0)^{(k-1)N}\log L(0)\right)_{\ge 0} 
  + \left(\bar{L}(0)^{-(k-1)\bar{N}}\log\bar{L}(0)\right)_{<0}. 
\end{gathered}
\]
\end{prop}

\begin{proof}
(\ref{eq-TW-eq}), (\ref{eq-TbarW-eq}), (\ref{eq-TWbar-eq}) 
and(\ref{eq-TbarWbar-eq}) are a direct consequence 
of the Sato equations for $W$ and $\bar{W}$.  
In order to derive (\ref{eq-XW-eq}) and (\ref{eq-XWbar-eq}), 
we consider the operator 
\[
  \calL_k = (L^N - \nu\log L)^k 
  = (\bar{L}^{-\bar{N}} - \nu\log\bar{L})^k. 
\]
This operator, just like $\calL$, can be expressed as 
\[
\begin{aligned}
  \calL_k 
  &= \left((L^N - \nu\log L)^k\right)_{\ge 0} 
  + \left((\bar{L}^{-\bar{N}} - \nu\log\bar{L})^k\right)_{<0}\\
  &= B_{kN} + \bar{B}_{k\bar{N}} - \nu k C_k + O(\nu^2), 
\end{aligned}
\]
where 
\[
  C_k = \left(L^{(k-1)N}\log L\right)_{\ge 0} 
        + \left(\bar{L}^{-(k-1)\bar{N}}\log\bar{L}\right)_{<0}. 
\]
This operator and the dressing operators satisfy 
the operator identities 
\[
\begin{aligned}
  \calL_kW &= W(\Lambda^N - \nu\log\Lambda)^k,\\
  \calL_k\bar{W} &= \bar{W}(\Lambda^{-\bar{N}} - \nu\log\Lambda). 
\end{aligned}
\]
By the Sato equations, the left hand side 
of the first identity can be expressed as 
\[
\begin{aligned}
  \calL_kW &= B_{kN}W + \bar{B}_{k\bar{N}}W - k\nu C_kW + O(\nu^2)\\
  &= \hbar\frac{\rd W}{\rd t_{kN}} 
     + \hbar\frac{\rd W}{\rd\bar{t}_{k\bar{N}}} 
     + W\Lambda^{kN} - k\nu C_kW + O(\nu^2)\\
  &= \hbar\nu\frac{\rd W}{\rd X_k} 
     + W\Lambda^{kN} - k\nu C_kW + O(\nu^2), 
\end{aligned}
\]
whereas the right hand side can be expanded as 
\[
  W(\Lambda^N - \nu\log\Lambda)^k 
  = W\Lambda^{kN} - k\nu W\Lambda^{(k-1)N}\log\Lambda + O(\nu^2). 
\]
Extracting the coefficients of $\nu$ in these expressions, 
we obtain (\ref{eq-XW-eq}). In the much same way, 
we can derive (\ref{eq-XWbar-eq}).  
\end{proof}

\begin{cor}
$L(0)$ and $\bar{L}(0)$ satisfy the Lax equations 
\begin{align}
  \hbar\frac{\rd L(0)}{\rd T_k} &= [B_k(0),L(0)] 
    \quad \text{for $k = 1,2,\ldots$}, \\
  \hbar\frac{\rd L(0)}{\rd\bar{T}_k} &= [\bar{B}_k(0),L(0)] 
    \quad \text{for $k \not\equiv 0 \mod \bar{N}$}, \\
  \hbar\frac{\rd L(0)}{\rd X_k} &= [kC_k(0),L(0)] 
    \quad \text{for $k = 1,2,\ldots$} 
\end{align}
and
\begin{align}
  \hbar\frac{\rd\bar{L}(0)}{\rd T_k} &= [B_k(0),\bar{L}(0)] 
    \quad \text{for $k = 1,2,\ldots$}, \\
  \hbar\frac{\rd\bar{L}(0)}{\rd\bar{T}_k} &= [\bar{B}_k(0),\bar{L}(0)] 
    \quad \text{for $k \not\equiv 0 \mod \bar{N}$}, \\
  \hbar\frac{\rd\bar{L}(0)}{\rd X_k} &= [kC_k(0),\bar{L}(0)] 
    \quad \text{for $k = 1,2,\ldots$}. 
\end{align}
\end{cor}

Thus we can derive all Lax and Sato equations 
of the extended bigraded Toda hierarchy \cite{Carlet06} 
from the equivariant bigraded Toda hierarchy 
in the scaling limit as $\nu \to 0$.  

\begin{remark}
The soliton solutions of the 2D Toda hierarchy 
are given by tau functions of the same form 
as shown in (\ref{lKP-soliton}) 
with the exponential factors 
\[
  e_i = (p_i/q_i)^{s/\hbar} 
    \exp\left(\hbar^{-1}(\xi(\bst,p_i) - \xi(\bst,q_i) 
      + \xi(\bar{\bst},p_i^{-1}) - \xi(\bar{\bst},q_i^{-1}))\right). 
\]
This tau function takes the reduced form (\ref{eq-red-tau}) 
if the parameters $p_i,q_i$ satisfy the equations 
\beq
  p_i^N + p_i^{-\bar{N}} - \nu\log p_i 
  = q_i^N + q_i^{-\bar{N}} - \nu\log q_i, \quad i = 1,\ldots,M. 
\eeq
Thus we obtain soliton solutions of the equivariant 
bigraded Toda hierarchy.  The prescription 
of the scaling limit will be applicable 
to these soliton solutions as well 
(though in a technically more complicated way). 
\end{remark}

\end{document}